\documentclass[aps,prl,twocolumn,showpacs,reprint,superscriptaddress,longbibliography]{revtex4-2}

\usepackage{amsmath, amsfonts, amssymb, bm}
\usepackage{mathtools}
\usepackage[inline]{enumitem}
\usepackage{graphicx}
\usepackage{xcolor}
\usepackage{lineno}
\usepackage[breaklinks,hypertexnames=false]{hyperref}
\hypersetup{colorlinks,linkcolor={magenta},citecolor={blue},urlcolor={blue}}

\usepackage[capitalize]{cleveref}

\usepackage{glossaries}
\glsdisablehyper     % disable hyperreferencing acronyms
\newacronym{mbs}{MBS}{Majorana bound state}
\newacronym{qsh}{QSHI}{quantum spin Hall insulator}
\newacronym{zbp}{ZBCP}{zero-bias conductance peak}

\begin{document}

\title{Tunable Majorana corner modes by orbital-dependent exchange interaction in a two-dimensional topological superconductor}

\author{Bo Lu}
%\affiliation{\tianjin}

\author{Yiying Zhang}
%\affiliation{\tianjin}
	
\affiliation{Center for Joint Quantum Studies and Department of Physics,
	Tianjin University, Tianjin 300072, China}

\date{\today}
	%\linenumbers
\begin{abstract}
	We theoretically study the effect of orbital-dependent exchange field in the formation of second order topological superconductors. We demonstrate that changing the orbital difference can induce topological transition and the Majorana corner modes therein can be manipulated. We further propose to detect the corner modes via a normal probe terminal.  The conductance quantization is found to be robust to changes of the relevant system parameters.
\end{abstract}

\maketitle

\emph{Introduction.---} The topological insulator (TI) is a new phase of
matter featuring non-trivial gapped band structure in the bulk and metallic
states on the boundary~\cite{Hasan10,Qi11}. It is known that TI can be
characterized by topological invariants~\cite{FuL07,FuB07} instead of an
order parameter. Meanwhile, introducing symmetry-breaking order to TI\
systems may provide new routes to more exotic quantum states. For instance,
by doping magnetic impurities in TI, the exchange interaction breaks
time-reversal symmetry and an energy gap can be opened at the Dirac point of
surface states, allowing the formation of quantum anomalous Hall insulators~%
\cite{Xufeng15,Chang16,He18,Tokura2019}. Furthermore, by proximity effect to
a superconductor where pair potential is induced, the magnetic TI may become
a topological superconductor~\cite{Alicea_2012,Been13,sato16} with midgap
states at the boundary. These predicted midgap states behave like Majorana
fermions and owing to their non-Abelian statistics, they are promising
building-blocks for fault-tolerant quantum computations~\cite%
{Kitaev01,Kitaev03,Nayak08}.

Very recently, a new class of topological superconductors coined
"second-order topological superconductors" (SOTS)~\cite%
{Langbehn17,Zhu18,Khalaf18,Wang18,Yan18,Fan18,Nori18,Liu19,Klin18,Klin19,Xiaoyu19,Fulga19,Ali19,Zeng19,Loss19,RoB19,Zhang20,Roy20,Chuan20,Nagai20,Zhang201,Zhang202,Tao21,Frank21,Xin21} is proposed and attracts much attention. SOTS is a $2$-dimensional
superconducting system with topologically nontrivial $0$-dimensional
Majorana corner modes (MCMs). Such corner states emerge at domain walls due
to the induced superconducting or magnetic gap along adjacent boundaries and
have been actively sought after in proximity-modified 2D TI systems, e.g., a
2D TI in proximity to high temperature superconductors~\cite%
{Yan18,Fan18,Nori18} or to $s$-wave superconductors under an in-plane
magnetic field~\cite{Chuan20}.

It is noted that there are some subtle issues in TI-based SOTS by either
magnetic field or magnetic doping. In previous studies~\cite{Chuan20}, the
exchange coupling to the magnetization is assumed to be uniform while it may
have different weight for orbitals in actual TI materials. The
orbital-dependent exchange interactions can result in a drastic change of
edge states and therefore it is worth investigating and clarifying its
effect. Indeed, attention has been paid to the higher-order TI~ \cite{Niu20}%
. However, as far as we know, there is no parallel study in the
superconducting phase. Another issue is how to detect MCMs in SOTS. Besides
theoretical prediction, MCMs have not been observed in experiments and a
feasible probe design is still needed.

In this context, we study the effect of orbital-dependent exchange field in
a magnetic TI coupled to superconductors. It is found that the different
weight of exchange field can drive the topological phase transition from
SOTSs to topological trivial superconductors. Moreover, we propose to use a
metallic tip brought in contact with the corner of TI to measure the
tunneling spectroscopy. A robust quantized conductance is found and can be
served as a way to identify MCMs.

\emph{Model.---} We consider second order topological superconductors in a 2D magnetic TI by
introducing superconductivity via proximity effect. The Hamiltonian of the
system is $H=\frac{1}{2}\sum_{\mathbf{k}}\Psi _{\mathbf{k}}^{\dag }H_{%
	\mathbf{k}}\Psi _{\mathbf{k}}$ with $H_{\mathbf{k}}=H_{\mathbf{k,}0}+H_{%
	\mathbf{k,}s\left( d\right) }+H_{\mathbf{k,}M}$, $\Psi _{\mathbf{k}}=(\psi
_{a,\mathbf{k}\uparrow },\psi _{b,\mathbf{k}\uparrow },\psi _{a,\mathbf{k}%
	\downarrow },\psi _{b,\mathbf{k}\downarrow },\psi _{a,-\mathbf{k}\uparrow
}^{\dag },\psi _{b,-\mathbf{k}\uparrow }^{\dag },\psi _{a,-\mathbf{k}%
	\downarrow }^{\dag },\psi _{b,-\mathbf{k}\downarrow }^{\dag })^{T}$ and $%
\psi _{\sigma ,s}$ the field operators of $\sigma $-orbital and $s$-spin. $%
H_{0}$ describes the normal 2D TI or quantum spin Hall insulator
\begin{equation}
	H_{\mathbf{k,}0}=\left( m+\frac{t}{2}k_{x}^{2}+\frac{t}{2}k_{y}^{2}\right)
	\hat{\sigma}_{z}\hat{\tau}_{z}+\lambda \left[ k_{x}\hat{\sigma}_{x}\hat{s}%
	_{z}+k_{y}\hat{\sigma}_{y}\hat{\tau}_{z}\right] ,
\end{equation}%
where $\hat{\sigma}$, $\hat{s}$ and $\hat{\tau}$ are Pauli matrices in the
orbital, spin and Numbu space, respectively. $H_{\mathbf{k,}s\left( d\right)
}$ describes the induced superconducting pairing on TI. We assume $%
m=m_{0}-2t<0$ for topological insulator state. Here, we consider
conventional $s$-wave and unconventional $d$-wave pairings as follows:
\begin{eqnarray}
	H_{\mathbf{k,}s} &=&-\Delta _{s}\hat{s}_{y}\hat{\tau}_{y}, \\
	H_{\mathbf{k,}d} &=&-\frac{\Delta _{d}}{2}\left( k_{x}^{2}-k_{y}^{2}\right)
	\hat{s}_{y}\hat{\tau}_{y},
\end{eqnarray}%
with $\Delta _{s}$ and $\Delta _{d}$ being the amplitude of $s$-wave and $d$%
-wave pairing potential in the bulk, respectively. $H_{\mathbf{k,}M}$ is the
exchange interaction from either the doped magnetic impurities or external
in-plane magnetic field:%
\begin{eqnarray}
	H_{\mathbf{k,}M} &=&\left( 1-\alpha \right) m_{x}\hat{s}_{x}\hat{\tau}%
	_{z}+\left( 1-\alpha \right) m_{y}\hat{s}_{y}  \notag \\
	&&+\alpha m_{x}\hat{\sigma}_{z}\hat{s}_{x}\hat{\tau}_{z}+\alpha m_{y}\hat{%
		\sigma}_{z}\hat{s}_{y}.
\end{eqnarray}%
where $\alpha $ describes the difference between exchange fields of the two
orbitals. For $\alpha =0$, the exchange fields of two orbitals has the same
weight while $\alpha =1$ means that the anti-ferromagnetic order emerges.
The value of $\alpha $ runs from $0$ to $1$. $\left( m_{x},m_{y}\right)
=M_{0}(\cos \theta ,\sin \theta )$ describe the in-plane magnetic field with
$M_{0}$ and $\theta $ being the magnitude and direction, respectively.
Unlike in-plane field, the out--of-plane magnetic field does not break the
mirror symmetry and thus is irrelevant of producing mass term \cite{Nori18}.

The effective Hamiltonian $\tilde{H}=\tilde{H}_{0}+\tilde{H}_{s\left(
	d\right) }+\tilde{H}_{M}$ on edge $l\in \left( \text{I,II,III,IV}\right) $
can be obtained in the standard way \cite{shenbook}, and they are%
\begin{equation}
	\tilde{H}_{0,l}=i\lambda \partial _{l}\hat{s}_{z},
\end{equation}%
\begin{equation}
	\tilde{H}_{s,l}=-\tilde{\Delta}_{s}\hat{s}_{y}\hat{\tau}_{y},  \label{e1}
\end{equation}%
\begin{equation}
	\tilde{H}_{d,l}=\left( -1\right) ^{l-1}\tilde{\Delta}_{d}\hat{s}_{y}\hat{\tau%
	}_{y},  \label{e3}
\end{equation}%
\begin{eqnarray}
	\tilde{H}_{M,l} &=&\alpha m_{x}\xi _{l}\hat{s}_{x}\hat{\tau}_{z}+\alpha
	m_{y}\xi _{l}\hat{s}_{y}  \notag \\
	&&+\left( 1-\alpha \right) \zeta _{l}m_{x}\hat{s}_{x}\hat{\tau}_{z}+\left(
	1-\alpha \right) \zeta _{l}m_{y}\hat{s}_{y},
\end{eqnarray}%
with $\xi _{l}=\left( 1,0,1,0\right) $, $\zeta _{l}=\left( 0,1,0,1\right) $,
and $\partial _{l}=\left( \partial _{y},-\partial _{x},-\partial
_{y},\partial _{x}\right) $ for I,II,III and IV edges, respectively. $\tilde{%
	\Delta}_{s}$ ($\tilde{\Delta}_{d}$) is the effective edge pair potential for
$s$-wave ($d$-wave) pairing and is given by $\tilde{\Delta}_{s}=\Delta _{s}$
($\tilde{\Delta}_{d}=m\Delta _{d}/t$). It can be seen that apart from $%
\alpha =0$ and $1$, all edges have non-zero magnetic mass term. To study the
MCMs in real space, we can transform the Hamiltonian $H$ under Fourier
transformation $\Psi _{\mathbf{k}}=N^{-1/2}\sum\nolimits_{_{\left(
		i,j\right) }}e^{i\mathbf{k}x_{\left( i,j\right) }}\Psi _{i,j}$ and obtain
the $H=\frac{1}{2}(\hat{H}_{0}+\hat{H}_{s\left( d\right) }+\hat{H}_{M})$ in
the real space as follows
\begin{eqnarray}
	\hat{H}_{0} &=&\sum_{i,j}\left( m+2t\right) \Psi _{i,j}^{\dag }\hat{\sigma}%
	_{z}\hat{\tau}_{z}\Psi _{i,j}  \notag \\
	&&+\sum_{i,j}\Psi _{i,j}^{\dag }\left[ -\frac{t}{2}\hat{\sigma}_{z}\hat{\tau}%
	_{z}-\frac{\lambda }{2i}\hat{\sigma}_{x}\hat{s}_{z}\right] \Psi _{i+1,j}+H.c.
	\notag \\
	&&+\sum_{i,j}\Psi _{i,j}^{\dag }\left[ -\frac{t}{2}\hat{\sigma}_{z}\hat{\tau}%
	_{z}-\frac{\lambda }{2i}\hat{\sigma}_{y}\hat{\tau}_{z}\right] \Psi
	_{i,j+1}+H.c.,
\end{eqnarray}%
\begin{equation}
	\hat{H}_{s}=-\sum_{i,j}\Delta _{s}\Psi _{i,j}^{\dag }\hat{s}_{y}\hat{\tau}%
	_{y}\Psi _{i,j},
\end{equation}%
\begin{eqnarray}
	\hat{H}_{d} &=&-\sum_{i,j}\frac{\Delta _{d}}{2}[\Psi _{i+1,j}^{\dag }\hat{s}%
	_{y}\hat{\tau}_{y}\Psi _{i,j}+\Psi _{i,j}^{\dag }\hat{s}_{y}\hat{\tau}%
	_{y}\Psi _{i+1,j}  \notag \\
	&&-\Psi _{i,j+1}^{\dag }\hat{s}_{y}\hat{\tau}_{y}\Psi _{i,j}-\Psi
	_{i,j}^{\dag }\hat{s}_{y}\hat{\tau}_{y}\Psi _{i,j+1}],
\end{eqnarray}%
\begin{eqnarray}
	\hat{H}_{M} &=&\sum_{i,j}\left( 1-\alpha \right) \Psi _{i,j}^{\dag }\left[
	m_{x}\hat{s}_{x}\hat{\tau}_{z}+m_{y}\hat{s}_{y}\right] \Psi _{i,j}  \notag \\
	&&+\sum_{i,j}\alpha \Psi _{i,j}^{\dag }\left[ m_{x}\hat{\sigma}_{z}\hat{s}%
	_{x}\hat{\tau}_{z}+\alpha m_{y}\hat{\sigma}_{z}\hat{s}_{y}\right] \Psi
	_{i,j}.
\end{eqnarray}%
We define $\hat{H}$ as $H=\frac{1}{2}\sum_{\mathbf{i,j}}\Psi _{\mathbf{i}%
}^{\dag }\hat{H}_{\mathbf{i},\mathbf{j}}\Psi _{\mathbf{j}}$ where $\mathbf{i=%
}\left( i,j\right) $ denotes the lattice site. The Green's function can be
obtained as
\begin{equation}
	\hat{g}^{r}\left( \varepsilon \right) =\frac{1}{\varepsilon +i0^{+}-\hat{H}}.
\end{equation}%
The local density of states (LDOS) at energy $\varepsilon $ on site $\mathbf{%
	i}$ can be solved from $\hat{g}^{r}\left( \varepsilon \right) $:
\begin{equation}
	\rho _{\mathbf{i}}\left( \varepsilon \right) =-\frac{1}{\pi }\text{Im }\text{%
		Tr}[\hat{g}_{\mathbf{i,i}}^{r}\left( \varepsilon \right) _{11}+\hat{g}_{%
		\mathbf{i,i}}^{r}\left( \varepsilon \right) _{22}].
\end{equation}%
For comparison, we can define the normalized LDOS $\tilde{\rho}_{\mathbf{i}%
}\left( \varepsilon \right) =\rho _{\mathbf{i}}\left( \varepsilon \right) /%
\left[ \sum_{\mathbf{i}}\rho _{\mathbf{i}}\left( \varepsilon \right) \right]
$. We choose a square 2D TI denoted by $N\times N$, where $N$ count the
equal atom numbers along $x$ and $y$ directions.

\begin{figure*}[tbp]
	\includegraphics[width=0.8\textwidth]{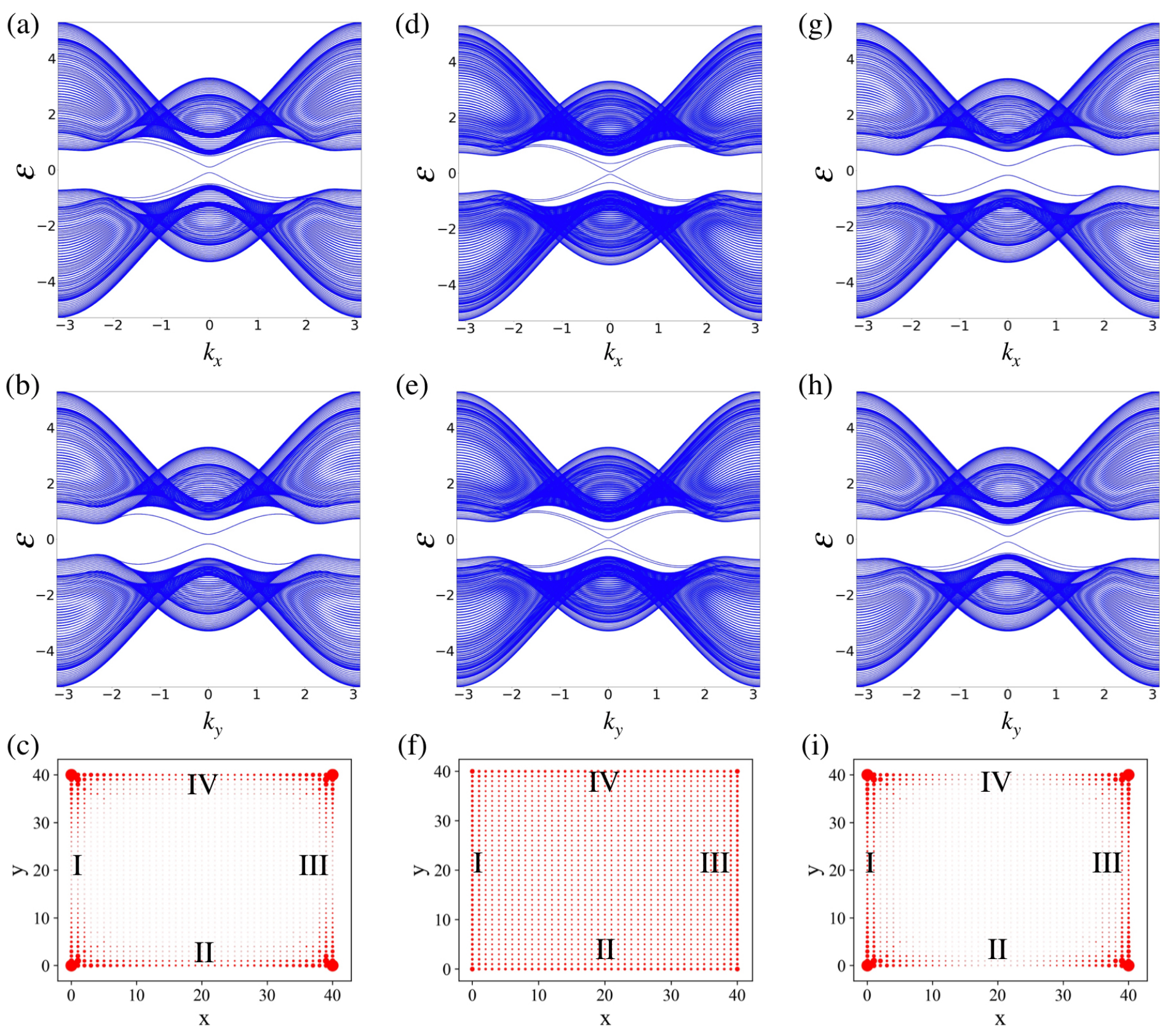}
	\caption{Band structure of a ribbon geometry along (a) $x$-direction and
		(b) $y$-direction with $s$-wave pairings and $\protect\alpha =0$. 60
		lattices is chosen along the direction with open boundary condition. (c) The
		normalized local density of states at $\protect\varepsilon=0$ for the same
		sample with lattices $N\times N=41\times 41$ and open boundary condition. In
		(d)-(f) $\protect\alpha =0.5$ and in (g)-(i) $\protect\alpha =1.$ Other
		parameters are the same for all panels: $m_{0}=1$, $t=2$, $\protect\lambda %
		=1 $, $\Delta _{s}=0.2 $, $m_{x}=0.3$ and $m_y=0$.}
	\label{fig:02}
\end{figure*}

\begin{figure*}[tbp]
	\includegraphics[width=0.8\textwidth]{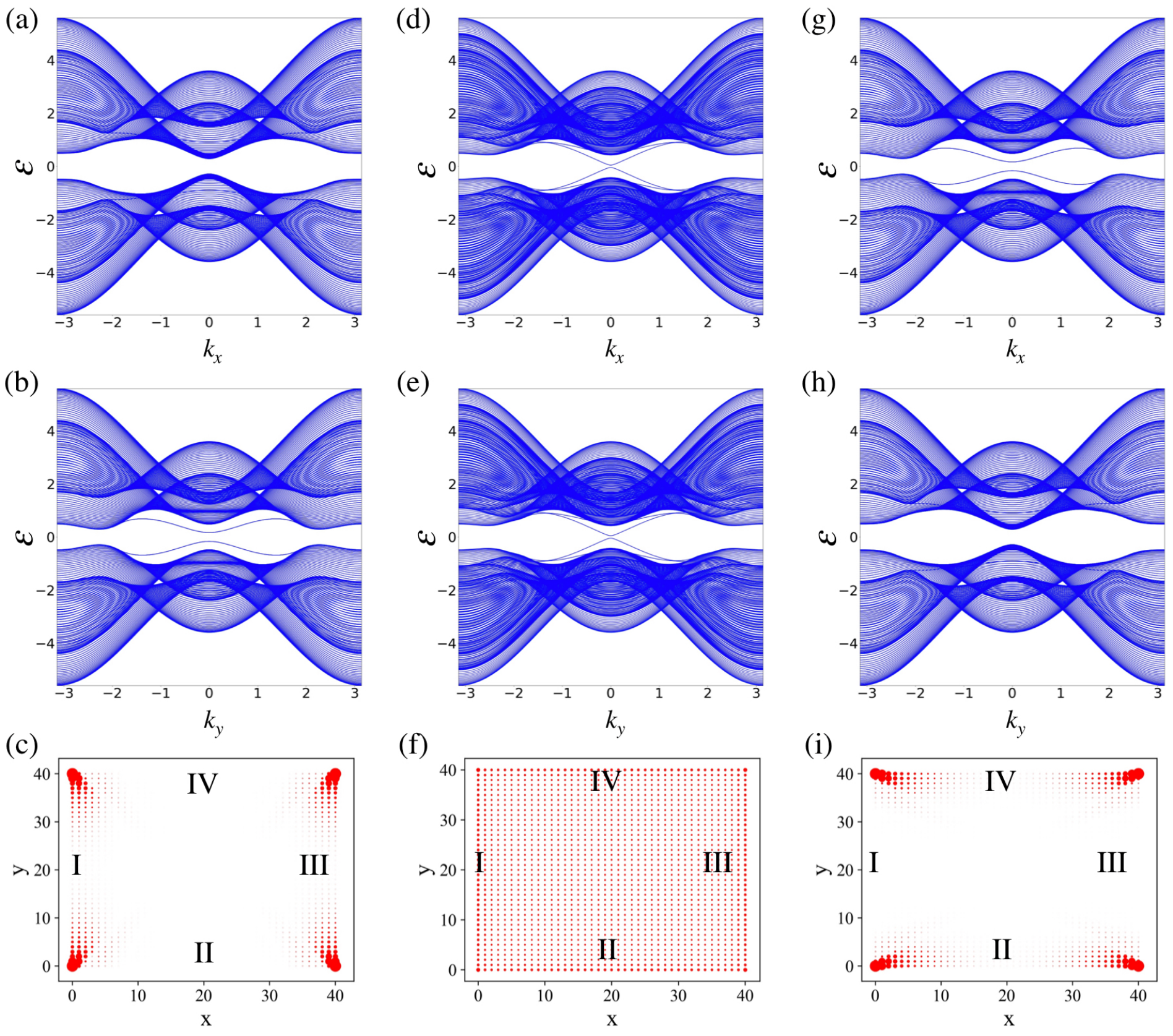}
	\caption{Band structure and normalized local density of states at $\protect%
		\varepsilon=0$ for $d$-wave pairings. $\protect\alpha =0$ for left panels, $%
		\protect\alpha =0.5$ for middle panels, and $\protect\alpha =1$ for right
		panels. Other parameters are the same as \cref{fig:02} except for $\Delta
		_{d}=0.2$ and $m_{x}=0.6$. }
	\label{fig:03}
\end{figure*}

\emph{Majorana corner modes.---} We first show the energy spectrum of 2D
magnetic TI in proximity to a superconductor. The aim for the band
calculation is to testify the gap opening by Dirac mass in Eqs. \ref{e1}-\ref{e3}. We use a ribbon geometry with open boundary conditions along the $x$ direction or $y$ direction and generate the band structure of $\hat{H}_{\mathbf{i},\mathbf{j}}$ using exact diagonalization. The lattices of the ribbon width is fixed at $N = 60$.  To obtain the LDOSs, we use a sample with lattices $N\times N=41\times 41$ and open boundary condition.

Figure \ref{fig:02} plots band structure and local density of states in the $%
s$-wave pairing state. The degenerate energy bands split due to non-zero
in-plane magnetic field. When $\alpha =0$, two orbitals experience the same
exchange interaction, and we recover the known results of MCMs ~\cite%
{Chuan20} in \cref{fig:02}(c). However, we find no MCMs for $\alpha =0.5$
with same band parameters as shown in \cref{fig:02}(f). When $\alpha =1$,
the MCMs reappear at four corners (see \cref{fig:02} (i)) though the two
orbitals now have opposite exchange splittings. This indicates that the
orbital difference $\alpha $ is a critical parameter which can induce the
topological phase transition in SOTS. Then we calculate the same sample by
replacing $s$-wave paring state by $d$-wave one as shown in \cref{fig:03}.
We observe the topological phase transition by $\alpha $ and for large $m_x $%
. It is necessary to point out that by measuring the local density of states
alone can not identify the number of MCMs located at the four corners of the
sample.

\begin{figure}[t]
	\centering{}\includegraphics[width=0.48\textwidth]{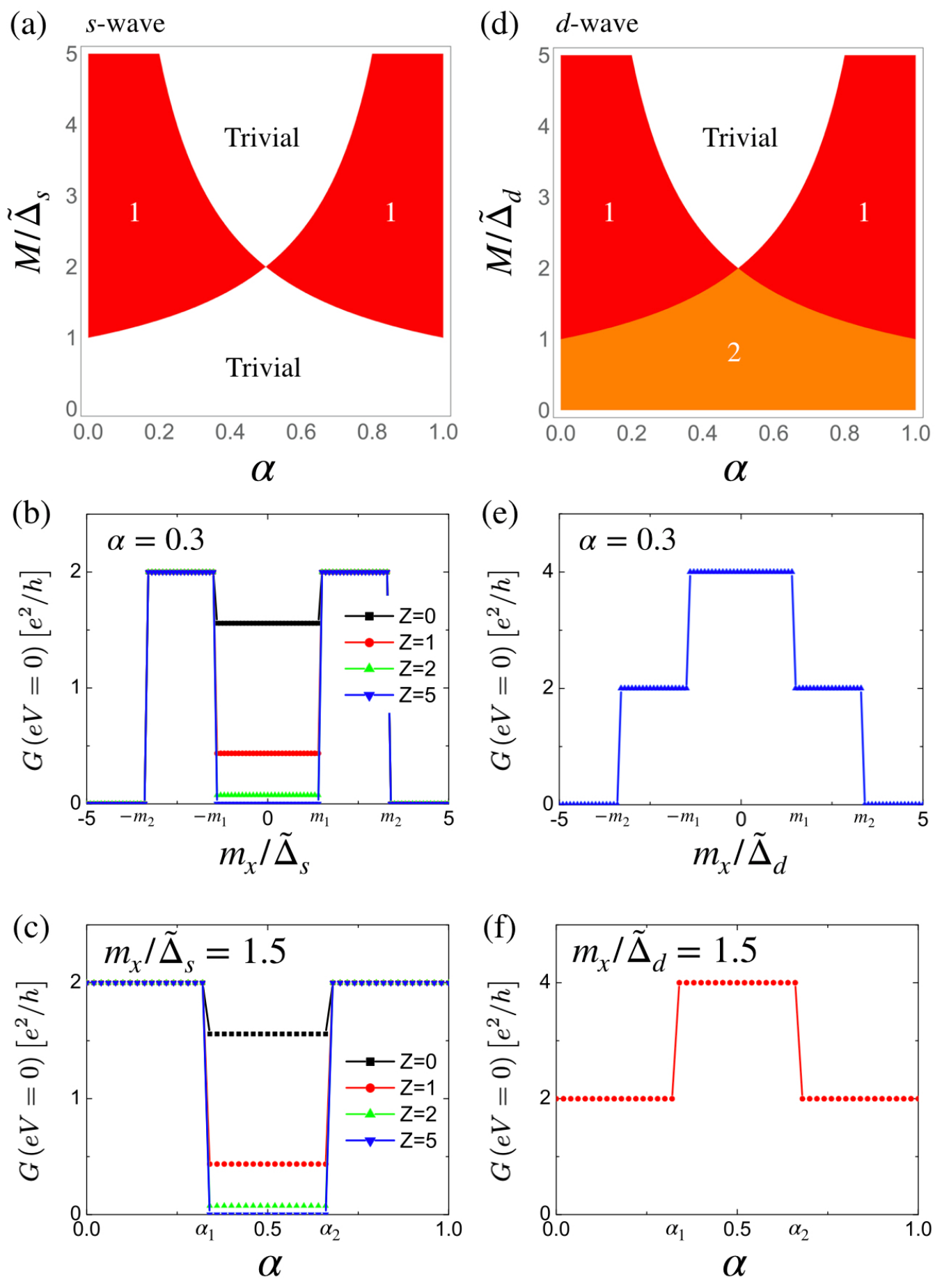}
	\caption{(a)(d) Phase diagram of superconductivity in the $\protect\alpha $-$%
		M/ \tilde{\Delta}_{s\left( d\right) }$ plane. The white region is
		topologically trivial. The region 1 (red) indicates the phase with a single
		MCM at each corner. The region 2 (orange) indicates the phase with a pair of
		MCMs at each corner. (b)(c) are the zero bias conductance for $s$-wave
		pairing state while (e) and (f) for $d$-wave case.}
	\label{fig:04}
\end{figure}

\emph{Topological analysis.---} Let us investigate the condition for MCMs based on effective Hamiltonian on
edges. We choose two specific adjacent edges: I and IV. Making a unitary
transformation
\begin{equation}
	U=\frac{1}{\sqrt{2}}\left[
	\begin{array}{cccc}
		1 & 0 & e^{-i\theta } & 0 \\
		0 & 1 & 0 & -e^{i\theta } \\
		0 & e^{-i\theta } & 0 & 1 \\
		e^{i\theta } & 0 & -1 & 0%
	\end{array}%
	\right] ,
\end{equation}%
we find that the I edge can be described by the following Hamiltonian $%
\breve{H}_{\text{I}}=U\tilde{H}_{\text{I}}U^{\dag }=h_{\text{I},1}\oplus h_{%
	\text{I},2}$, with
\begin{equation}
	h_{\text{I},1}=\left[
	\begin{array}{cc}
		i\lambda \partial _{y} & e^{-i\theta }\left[ \alpha M\mp \tilde{\Delta}%
		_{s\left( d\right) }\right] \\
		e^{i\theta }\left[ \alpha M\mp \tilde{\Delta}_{s\left( d\right) }\right] &
		-i\lambda \partial _{y}%
	\end{array}%
	\right] ,
\end{equation}%
\begin{equation}
	h_{\text{I},2}=\left[
	\begin{array}{cc}
		-i\lambda \partial _{y} & e^{-i\theta }\left[ \alpha M\pm \tilde{\Delta}%
		_{s\left( d\right) }\right] \\
		e^{i\theta }\left[ \alpha M\pm \tilde{\Delta}_{s\left( d\right) }\right] &
		i\lambda \partial _{y}%
	\end{array}%
	\right] .
\end{equation}%
and the IV edge is transformed into $\breve{H}_{\text{IV}}=U\tilde{H}_{\text{%
		IV}}U^{\dag }=h_{\text{IV},1}\oplus h_{\text{IV},2}$, with%
\begin{eqnarray}
	h_{\text{IV},1} &=&\left[
	\begin{array}{cc}
		i\lambda \partial _{x} & e^{-i\theta }\Lambda _{1} \\
		e^{i\theta }\Lambda _{1} & -i\lambda \partial _{x}%
	\end{array}%
	\right] , \\
	h_{\text{IV},2} &=&\left[
	\begin{array}{cc}
		-i\lambda \partial _{x} & e^{-i\theta }\Lambda _{2} \\
		e^{i\theta }\Lambda _{2} & i\lambda \partial _{x}%
	\end{array}%
	\right] , \\
	\Lambda _{1} &=&\left( 1-\alpha \right) M-\tilde{\Delta}_{s\left( d\right) },
	\\
	\Lambda _{2} &=&\left( 1-\alpha \right) M+\tilde{\Delta}_{s\left( d\right) }.
\end{eqnarray}%
According to the Jackiw-Rebbi theory~\cite{JR}, the existence of MCMs for $s$%
-wave pair potential requires that
\begin{equation}
	\left( \alpha M/\tilde{\Delta}_{s}-1\right) \left[ \left( 1-\alpha \right) M/%
	\tilde{\Delta}_{s}-1\right] <0.  \label{ts}
\end{equation}%
It is noted that angle $\theta $ does not plays a role in Eq. \ref{ts}, thus
the topological condition only depends on the magnitude of magnetization $M$
and effective pair potential $\tilde{\Delta}_{s}$. The phase diagram is
shown in \cref{fig:04}(a) where the shaded region indicates the SOTS. Such
diagram agrees well with the numerical result in \cref{fig:02}. It can be
seen that under the condition $M>\tilde{\Delta}_{s}$, the system undergoes
topological phase transition by increasing $\alpha $ from $0$ to $1$.
Moreover, the system stays in the trivial states for an intermediate value
of $\alpha =0.5$ regardless of $M$.

For $d$-wave pair potential, it can be verified that pairs of MCMs can be
found under the condition $d_{1}<0$ and $d_{2}<0$ where $d_{1}$ and $d_{2}$
are given by%
\begin{eqnarray}
	d_{1} &=&\left( \alpha M/\tilde{\Delta}_{d}+1\right) \left[ \left( 1-\alpha
	\right) M/\tilde{\Delta}_{d}-1\right] , \\
	d_{2} &=&\left( \alpha M/\tilde{\Delta}_{d}-1\right) \left[ \left( 1-\alpha
	\right) M/\tilde{\Delta}_{d}+1\right] .
\end{eqnarray}%
When $d_{1}$ and $d_{2}$ satisfy $d_{1}d_{2}<0$, a single MCM exists. For $d$%
-wave pairings, the pair potential undergoes a sign change for adjacent
edges, which provides additional sign change of the mass
term leading to pairs of MCMs. It is also interesting to note that the
phase diagram of $d$-wave pairing (see \cref{fig:04}(d)) resembles the phase
diagram of $s$-wave case. However, pairs of MCMs form for $M<$ $\tilde{\Delta%
}_{d}$ regardless of $\alpha $ due to the unconventional $d$-wave pairing.

\emph{Experimental signatures.---} To explore the experimental signatures of the MCMs, we propose to use a
normal probe terminal, such as an STM tip, coupled to the corner of the
SOTS. Specifically, the semi-infinite normal probe on the $x$-axis is placed
at the corner between I and IV edges. We set the origin at the corner and
the Hamiltonian $H_{N}$ of the probe is
\begin{equation}
	H_{N}=\left[ -\frac{\partial _{x}^{2}}{2\tilde{m}}-\mu _{N}+U\delta \left(
	x+0^{+}\right) \right] \hat{\tau}_{z},
\end{equation}
where $\tilde{m}$, $\mu _{N}$ and $U$ are the effective mass, chemical
potential and the barrier parameter between the probe and the 2D TI,
respectively. For simplicity, we only consider $x$-axis magnetization. Using
the similar method \cite{Modak12,Soori13,Soori20}, one can obtain the
boundary condition which connects the wave functions $\Psi _{N}$ of the
normal probe, $\Psi _{\text{I}}$ and $\Psi _{\text{IV}}$ of the edge I and
IV at the corner as follows
\begin{eqnarray}
	\Psi _{N} &=&\chi \left( \Psi _{\text{I}}+\Psi _{\text{IV}}\right) , \\
	\chi \left( \partial _{x}\Psi _{N}+Zk\Psi _{N}\right) &=&i\lambda \tilde{m}
	\hat{s}_{z}\hat{\tau}_{z}\left( \Psi _{\text{I}}-\Psi _{\text{IV}}\right) .
\end{eqnarray}%
The parameter $Z=2\tilde{m}U/k$ ($k$: the Fermi wave vector) describes the
barrier strength, while the real and dimensionless number $\chi $ represents
the different microscopic details between the probe and 2D TI, such as the
hopping integrals in the underlying lattice model. The scattering wave
function for the normal probe is solved as
\begin{align}
	\Psi _{N\sigma }={}& \Phi _{\sigma }+b_{\uparrow \sigma }\hat{B}_{\uparrow
	}e^{-ikx}+b_{\downarrow \sigma }\hat{B}_{\downarrow }e^{-ikx}  \nonumber \\
	& +a_{\uparrow \sigma }\hat{A}_{\uparrow }e^{ikx}+a_{\downarrow \sigma }\hat{
		A}_{\downarrow }e^{ikx},
\end{align}%
for an incident electron with spin $\sigma =\uparrow (\downarrow )$ and wave
function $\Phi _{\uparrow }=(1,0,0,0)^{T}e^{iqx}$ [$\Phi _{\downarrow
}=(0,1,0,0)^{T}e^{iqx}$]. Under the wide band approximation, the wave vector
is $k=\sqrt{2m_{N}\mu _{N}}$ and the spinors are $\hat{B}_{\uparrow
}=(1,0,0,0)^{T}$, $\hat{B}_{\downarrow }=(0,1,0,0)^{T}$, $\hat{A}_{\uparrow
}=(0,0,1,0)^{T}$, and $\hat{A}_{\downarrow }=(0,0,0,1)^{T}$. The normal
(Andreev) reflection amplitudes are $b_{\sigma ^{\prime }\sigma }$ ($%
a_{\sigma ^{\prime }\sigma }$) for an incoming electron of spin $\sigma $
scattered as an electron (hole) of spin $\sigma ^{\prime }$. We focus on
zero energy solution and thus the wave functions of edge I and IV are%
\begin{eqnarray}
	\Psi _{\text{IV,}\sigma } &=&c_{1\sigma }\Phi _{1}\left( p_{1}\right)
	e^{-\left\vert p_{1}\right\vert x}+c_{2\sigma }\Phi _{2}\left( p_{2}\right)
	e^{-\left\vert p_{2}\right\vert x}, \\
	\Psi _{\text{I,}\sigma } &=&d_{1\sigma }\Phi _{3}\left( p_{3}\right)
	e^{\left\vert p_{3}\right\vert y}+d_{2\sigma }\Phi _{4}\left( p_{4}\right)
	e^{\left\vert p_{4}\right\vert y}.
\end{eqnarray}%
with $\Phi _{1\left( 2\right) }\left( \kappa \right) =(\pm \nu ,\mp 1,\nu
,1)^{T}$, $\Phi _{3\left( 4\right) }\left( \kappa \right) =(\nu ,1,\mp \nu
,\pm 1)^{T}$ and $\nu =i\kappa /\left\vert \kappa \right\vert $. For $s$%
-wave pair potential, we have $\lambda p_{1\left( 2\right) }=m_{x}-\alpha
m_{x}\mp \tilde{\Delta}_{s}$, $\lambda p_{3\left( 4\right) }=\alpha m_{x}\pm
\tilde{\Delta}_{s}$ and they are $\lambda p_{1\left( 2\right) }=m_{x}-\alpha
m_{x}\mp \tilde{\Delta}_{d}$, $\lambda p_{3\left( 4\right) }=\alpha m_{x}\mp
\tilde{\Delta}_{d}$ for $d$-wave case. The differential conductance $G$ at
zero temperature is calculated using the formula~\cite{BTK}
\begin{equation}
	G=\frac{e^{2}}{h}\sum\nolimits_{\sigma \sigma ^{\prime }}\left[ \delta
	_{\sigma \sigma ^{\prime }}+\left\vert a_{\sigma \sigma ^{\prime
	}}\right\vert ^{2}-\left\vert b_{\sigma \sigma ^{\prime }}\right\vert ^{2} %
	\right] .  \label{eq13}
\end{equation}

Here, $a_{\sigma \sigma ^{\prime }}$ and $b_{\sigma \sigma ^{\prime }}$ are
reflection amplitudes at zero bias. For the numerical calculations we choose
$k_{N}/\tilde{m}\approx 1.57\times 10^{6}$ m/s, $\mu _{N}\approx 7.0$ eV
corresponding to copper, $\left\vert \lambda \right\vert \approx 5.5\times
10^{5}$ m/s and $\chi =1$. We note that the feature of zero-bias conductance
is not material dependent.

First, we display the result of zero bias conductance for $s$-wave pairing
state. For a fixed $\alpha $, the phase boundary clearly appears in the
conductance dependence on the magnetization $m_{x}$ in \cref{fig:04}(b). It
can be seen that the zero-bias conductance clearly showcases the celebrated
Majorana zero bias peak with quantized height $2e^{2}/h$~\cite%
{Sengupta01,Bolech07,Akhmerov09,Tanaka09,Law09} for $m_{1}<\left\vert
m_{x}\right\vert <m_{2}$. And as the quantized conductance appears, it
remains a plateau by altering the barrier $Z$. For $\left\vert
m_{x}\right\vert <m_{1}$, the system enters into non-topological phase and
the conductance can be greatly affected by $Z$. Moreover, the conductance
becomes almost $0$ for $\left\vert m_{x}\right\vert >m_{2}$ as the system is
no longer topological. This suppressed conductance can be explained by a
large opening gap due to magnetization where all waves on edges become
evanescent. In \cref{fig:04}(c) , we show how a quantum plateau of the
zero-biased conductance can be tuned out for a given $m_{x}$, by altering
band difference $\alpha $. The conductance becomes sensitive to the barrier $%
Z$ for $\alpha _{1}<\alpha <\alpha _{2}$ which reflects no MCM. The result
of conductance agrees well with the phase diagram as shown in \cref{fig:04}%
(a) and provides a support for experimental detection of MCMs by tunneling
spectroscopy.

We now turn to the $d$-wave case. At zero magnetization, it has been known
that there are pairs of MCMs. Our conductance result shows a $4e^{2}/h$
conductance peak as the presence of pairs of MCMs even in a weak exchange
field $\left\vert m_{x}\right\vert <m_{1}$ in \cref{fig:04}(e). We do note
specify the used value of $Z$ since this conductance is immune to its
variation. For $m_{1}<\left\vert m_{x}\right\vert <m_{2}$, we obtain a $%
2e^{2}/h$ conductance plateau similar to the $s$-wave case, indicating a
single MCM at each corner. And as $\left\vert m_{x}\right\vert $ is greater
than $m_{2}$, the system become topological trivial and gives rises to
suppressed conductance due to the magnetic gap. It shows that the
conductance quantization is robust in each topological phase as $\alpha $
changes for a given $m_{x}$, as shown in \cref{fig:04}(f). These results are
consistent with the phase diagram in \cref{fig:04}(d). We can see that the
tunneling spectroscopy is a useful way of not only providing the information
of the presence of MCMs, but also the number of them.

\emph{Conclusions.---} We have studied the orbital-dependent exchange field
effect on the formation of second order topological superconductors based on
two-dimensional topological insulators. We have considered both $s$-wave and
$d$-wave pairing states. The Majorana corner modes are shown to be dependent
on the orbital difference. For experimental realizations, we expect 2D topological insulator materials such as HgTe quantum wells. When proximate to superconductors, a proximity-induced superconducting gap can be induced \cite{Bocquillon17}. And when doping with Mn, the magnetization of HgTe shows orbital difference under an in-plane magnetic field \cite{LiuX2013}.  And an in-plane magnetic field has been successfully coulped to HgTe quantum wells in recent experiments \cite{Ren2019}. We finally propose an experiment to demonstrate that the
quantized zero-biased conductance indeed arises due to the Majorana corner
modes.

\emph{Acknowledgments.---} We acknowledge support from the National Natural
Science Foundation of China (project 11904257) and the Natural Science
Foundation of Tianjin (project 20JCQNJC01310).
				
\bibliography{SpinSC}

%apsrev4-2.bst 2019-01-14 (MD) hand-edited version of apsrev4-1.bst
%Control: key (0)
%Control: author (8) initials jnrlst
%Control: editor formatted (1) identically to author
%Control: production of article title (0) allowed
%Control: page (0) single
%Control: year (1) truncated
%Control: production of eprint (0) enabled
\begin{thebibliography}{54}%
\makeatletter
\providecommand \@ifxundefined [1]{%
 \@ifx{#1\undefined}
}%
\providecommand \@ifnum [1]{%
 \ifnum #1\expandafter \@firstoftwo
 \else \expandafter \@secondoftwo
 \fi
}%
\providecommand \@ifx [1]{%
 \ifx #1\expandafter \@firstoftwo
 \else \expandafter \@secondoftwo
 \fi
}%
\providecommand \natexlab [1]{#1}%
\providecommand \enquote  [1]{``#1''}%
\providecommand \bibnamefont  [1]{#1}%
\providecommand \bibfnamefont [1]{#1}%
\providecommand \citenamefont [1]{#1}%
\providecommand \href@noop [0]{\@secondoftwo}%
\providecommand \href [0]{\begingroup \@sanitize@url \@href}%
\providecommand \@href[1]{\@@startlink{#1}\@@href}%
\providecommand \@@href[1]{\endgroup#1\@@endlink}%
\providecommand \@sanitize@url [0]{\catcode `\\12\catcode `\$12\catcode
  `\&12\catcode `\#12\catcode `\^12\catcode `\_12\catcode `\%12\relax}%
\providecommand \@@startlink[1]{}%
\providecommand \@@endlink[0]{}%
\providecommand \url  [0]{\begingroup\@sanitize@url \@url }%
\providecommand \@url [1]{\endgroup\@href {#1}{\urlprefix }}%
\providecommand \urlprefix  [0]{URL }%
\providecommand \Eprint [0]{\href }%
\providecommand \doibase [0]{https://doi.org/}%
\providecommand \selectlanguage [0]{\@gobble}%
\providecommand \bibinfo  [0]{\@secondoftwo}%
\providecommand \bibfield  [0]{\@secondoftwo}%
\providecommand \translation [1]{[#1]}%
\providecommand \BibitemOpen [0]{}%
\providecommand \bibitemStop [0]{}%
\providecommand \bibitemNoStop [0]{.\EOS\space}%
\providecommand \EOS [0]{\spacefactor3000\relax}%
\providecommand \BibitemShut  [1]{\csname bibitem#1\endcsname}%
\let\auto@bib@innerbib\@empty
%</preamble>
\bibitem [{\citenamefont {Hasan}\ and\ \citenamefont {Kane}(2010)}]{Hasan10}%
  \BibitemOpen
  \bibfield  {author} {\bibinfo {author} {\bibfnamefont {M.~Z.}\ \bibnamefont
  {Hasan}}\ and\ \bibinfo {author} {\bibfnamefont {C.~L.}\ \bibnamefont
  {Kane}},\ }\bibfield  {title} {\bibinfo {title} {Colloquium: Topological
  insulators},\ }\href {https://doi.org/10.1103/RevModPhys.82.3045} {\bibfield
  {journal} {\bibinfo  {journal} {Rev. Mod. Phys.}\ }\textbf {\bibinfo {volume}
  {82}},\ \bibinfo {pages} {3045} (\bibinfo {year} {2010})}\BibitemShut
  {NoStop}%
\bibitem [{\citenamefont {Qi}\ and\ \citenamefont {Zhang}(2011)}]{Qi11}%
  \BibitemOpen
  \bibfield  {author} {\bibinfo {author} {\bibfnamefont {X.-L.}\ \bibnamefont
  {Qi}}\ and\ \bibinfo {author} {\bibfnamefont {S.-C.}\ \bibnamefont {Zhang}},\
  }\bibfield  {title} {\bibinfo {title} {Topological insulators and
  superconductors},\ }\href {https://doi.org/10.1103/RevModPhys.83.1057}
  {\bibfield  {journal} {\bibinfo  {journal} {Rev. Mod. Phys.}\ }\textbf
  {\bibinfo {volume} {83}},\ \bibinfo {pages} {1057} (\bibinfo {year}
  {2011})}\BibitemShut {NoStop}%
\bibitem [{\citenamefont {Fu}\ \emph {et~al.}(2007)\citenamefont {Fu},
  \citenamefont {Kane},\ and\ \citenamefont {Mele}}]{FuL07}%
  \BibitemOpen
  \bibfield  {author} {\bibinfo {author} {\bibfnamefont {L.}~\bibnamefont
  {Fu}}, \bibinfo {author} {\bibfnamefont {C.~L.}\ \bibnamefont {Kane}},\ and\
  \bibinfo {author} {\bibfnamefont {E.~J.}\ \bibnamefont {Mele}},\ }\bibfield
  {title} {\bibinfo {title} {Topological insulators in three dimensions},\
  }\href {https://doi.org/10.1103/PhysRevLett.98.106803} {\bibfield  {journal}
  {\bibinfo  {journal} {Phys. Rev. Lett.}\ }\textbf {\bibinfo {volume} {98}},\
  \bibinfo {pages} {106803} (\bibinfo {year} {2007})}\BibitemShut {NoStop}%
\bibitem [{\citenamefont {Fu}\ and\ \citenamefont {Kane}(2007)}]{FuB07}%
  \BibitemOpen
  \bibfield  {author} {\bibinfo {author} {\bibfnamefont {L.}~\bibnamefont
  {Fu}}\ and\ \bibinfo {author} {\bibfnamefont {C.~L.}\ \bibnamefont {Kane}},\
  }\bibfield  {title} {\bibinfo {title} {Topological insulators with inversion
  symmetry},\ }\href {https://doi.org/10.1103/PhysRevB.76.045302} {\bibfield
  {journal} {\bibinfo  {journal} {Phys. Rev. B}\ }\textbf {\bibinfo {volume}
  {76}},\ \bibinfo {pages} {045302} (\bibinfo {year} {2007})}\BibitemShut
  {NoStop}%
\bibitem [{\citenamefont {Kou}\ \emph {et~al.}(2015)\citenamefont {Kou},
  \citenamefont {Fan}, \citenamefont {Lang}, \citenamefont {Upadhyaya},\ and\
  \citenamefont {Wang}}]{Xufeng15}%
  \BibitemOpen
  \bibfield  {author} {\bibinfo {author} {\bibfnamefont {X.}~\bibnamefont
  {Kou}}, \bibinfo {author} {\bibfnamefont {Y.}~\bibnamefont {Fan}}, \bibinfo
  {author} {\bibfnamefont {M.}~\bibnamefont {Lang}}, \bibinfo {author}
  {\bibfnamefont {P.}~\bibnamefont {Upadhyaya}},\ and\ \bibinfo {author}
  {\bibfnamefont {K.~L.}\ \bibnamefont {Wang}},\ }\bibfield  {title} {\bibinfo
  {title} {Magnetic topological insulators and quantum anomalous hall effect},\
  }\href {https://doi.org/https://doi.org/10.1016/j.ssc.2014.10.022} {\bibfield
   {journal} {\bibinfo  {journal} {Solid State Communications}\ }\textbf
  {\bibinfo {volume} {215-216}},\ \bibinfo {pages} {34} (\bibinfo {year}
  {2015})}\BibitemShut {NoStop}%
\bibitem [{\citenamefont {Chang}\ and\ \citenamefont {Li}(2016)}]{Chang16}%
  \BibitemOpen
  \bibfield  {author} {\bibinfo {author} {\bibfnamefont {C.-Z.}\ \bibnamefont
  {Chang}}\ and\ \bibinfo {author} {\bibfnamefont {M.}~\bibnamefont {Li}},\
  }\bibfield  {title} {\bibinfo {title} {Quantum anomalous hall effect in
  time-reversal-symmetry breaking topological insulators},\ }\href
  {https://doi.org/10.1088/0953-8984/28/12/123002} {\bibfield  {journal}
  {\bibinfo  {journal} {Journal of Physics: Condensed Matter}\ }\textbf
  {\bibinfo {volume} {28}},\ \bibinfo {pages} {123002} (\bibinfo {year}
  {2016})}\BibitemShut {NoStop}%
\bibitem [{\citenamefont {He}\ \emph {et~al.}(2018)\citenamefont {He},
  \citenamefont {Wang},\ and\ \citenamefont {Xue}}]{He18}%
  \BibitemOpen
  \bibfield  {author} {\bibinfo {author} {\bibfnamefont {K.}~\bibnamefont
  {He}}, \bibinfo {author} {\bibfnamefont {Y.}~\bibnamefont {Wang}},\ and\
  \bibinfo {author} {\bibfnamefont {Q.-K.}\ \bibnamefont {Xue}},\ }\bibfield
  {title} {\bibinfo {title} {Topological materials: Quantum anomalous hall
  system},\ }\href {https://doi.org/10.1146/annurev-conmatphys-033117-054144}
  {\bibfield  {journal} {\bibinfo  {journal} {Annual Review of Condensed Matter
  Physics}\ }\textbf {\bibinfo {volume} {9}},\ \bibinfo {pages} {329} (\bibinfo
  {year} {2018})}\BibitemShut {NoStop}%
\bibitem [{\citenamefont {Tokura}\ \emph {et~al.}(2019)\citenamefont {Tokura},
  \citenamefont {Yasuda},\ and\ \citenamefont {Tsukazaki}}]{Tokura2019}%
  \BibitemOpen
  \bibfield  {author} {\bibinfo {author} {\bibfnamefont {Y.}~\bibnamefont
  {Tokura}}, \bibinfo {author} {\bibfnamefont {K.}~\bibnamefont {Yasuda}},\
  and\ \bibinfo {author} {\bibfnamefont {A.}~\bibnamefont {Tsukazaki}},\
  }\bibfield  {title} {\bibinfo {title} {Magnetic topological insulators},\
  }\href {https://doi.org/10.1038/s42254-018-0011-5} {\bibfield  {journal}
  {\bibinfo  {journal} {Nature Reviews Physics}\ }\textbf {\bibinfo {volume}
  {1}},\ \bibinfo {pages} {126} (\bibinfo {year} {2019})}\BibitemShut {NoStop}%
\bibitem [{\citenamefont {Alicea}(2012)}]{Alicea_2012}%
  \BibitemOpen
  \bibfield  {author} {\bibinfo {author} {\bibfnamefont {J.}~\bibnamefont
  {Alicea}},\ }\bibfield  {title} {\bibinfo {title} {New directions in the
  pursuit of majorana fermions in solid state systems},\ }\href
  {https://doi.org/10.1088/0034-4885/75/7/076501} {\bibfield  {journal}
  {\bibinfo  {journal} {Reports on Progress in Physics}\ }\textbf {\bibinfo
  {volume} {75}},\ \bibinfo {pages} {076501} (\bibinfo {year}
  {2012})}\BibitemShut {NoStop}%
\bibitem [{\citenamefont {Beenakker}(2013)}]{Been13}%
  \BibitemOpen
  \bibfield  {author} {\bibinfo {author} {\bibfnamefont {C.}~\bibnamefont
  {Beenakker}},\ }\bibfield  {title} {\bibinfo {title} {Search for majorana
  fermions in superconductors},\ }\href
  {https://doi.org/10.1146/annurev-conmatphys-030212-184337} {\bibfield
  {journal} {\bibinfo  {journal} {Annual Review of Condensed Matter Physics}\
  }\textbf {\bibinfo {volume} {4}},\ \bibinfo {pages} {113} (\bibinfo {year}
  {2013})}\BibitemShut {NoStop}%
\bibitem [{\citenamefont {Sato}\ and\ \citenamefont {Fujimoto}(2016)}]{sato16}%
  \BibitemOpen
  \bibfield  {author} {\bibinfo {author} {\bibfnamefont {M.}~\bibnamefont
  {Sato}}\ and\ \bibinfo {author} {\bibfnamefont {S.}~\bibnamefont
  {Fujimoto}},\ }\bibfield  {title} {\bibinfo {title} {Majorana fermions and
  topology in superconductors},\ }\href
  {https://doi.org/10.7566/JPSJ.85.072001} {\bibfield  {journal} {\bibinfo
  {journal} {Journal of the Physical Society of Japan}\ }\textbf {\bibinfo
  {volume} {85}},\ \bibinfo {pages} {072001} (\bibinfo {year}
  {2016})}\BibitemShut {NoStop}%
\bibitem [{\citenamefont {Kitaev}(2001)}]{Kitaev01}%
  \BibitemOpen
  \bibfield  {author} {\bibinfo {author} {\bibfnamefont {A.}~\bibnamefont
  {Kitaev}},\ }\bibfield  {title} {\bibinfo {title} {Unpaired majorana fermions
  in quantum wires},\ }\href {https://doi.org/10.1070/1063-7869/44/10s/s29}
  {\bibfield  {journal} {\bibinfo  {journal} {Physics-Uspekhi}\ }\textbf
  {\bibinfo {volume} {44}},\ \bibinfo {pages} {131} (\bibinfo {year}
  {2001})}\BibitemShut {NoStop}%
\bibitem [{\citenamefont {Kitaev}(2003)}]{Kitaev03}%
  \BibitemOpen
  \bibfield  {author} {\bibinfo {author} {\bibfnamefont {A.}~\bibnamefont
  {Kitaev}},\ }\bibfield  {title} {\bibinfo {title} {Fault-tolerant quantum
  computation by anyons},\ }\href
  {https://doi.org/10.1016/S0003-4916(02)00018-0} {\bibfield  {journal}
  {\bibinfo  {journal} {Annals of Physics}\ }\textbf {\bibinfo {volume}
  {303}},\ \bibinfo {pages} {2} (\bibinfo {year} {2003})}\BibitemShut {NoStop}%
\bibitem [{\citenamefont {Nayak}\ \emph {et~al.}(2008)\citenamefont {Nayak},
  \citenamefont {Simon}, \citenamefont {Stern}, \citenamefont {Freedman},\ and\
  \citenamefont {Das~Sarma}}]{Nayak08}%
  \BibitemOpen
  \bibfield  {author} {\bibinfo {author} {\bibfnamefont {C.}~\bibnamefont
  {Nayak}}, \bibinfo {author} {\bibfnamefont {S.~H.}\ \bibnamefont {Simon}},
  \bibinfo {author} {\bibfnamefont {A.}~\bibnamefont {Stern}}, \bibinfo
  {author} {\bibfnamefont {M.}~\bibnamefont {Freedman}},\ and\ \bibinfo
  {author} {\bibfnamefont {S.}~\bibnamefont {Das~Sarma}},\ }\bibfield  {title}
  {\bibinfo {title} {Non-abelian anyons and topological quantum computation},\
  }\href {https://doi.org/10.1103/RevModPhys.80.1083} {\bibfield  {journal}
  {\bibinfo  {journal} {Rev. Mod. Phys.}\ }\textbf {\bibinfo {volume} {80}},\
  \bibinfo {pages} {1083} (\bibinfo {year} {2008})}\BibitemShut {NoStop}%
\bibitem [{\citenamefont {Langbehn}\ \emph {et~al.}(2017)\citenamefont
  {Langbehn}, \citenamefont {Peng}, \citenamefont {Trifunovic}, \citenamefont
  {von Oppen},\ and\ \citenamefont {Brouwer}}]{Langbehn17}%
  \BibitemOpen
  \bibfield  {author} {\bibinfo {author} {\bibfnamefont {J.}~\bibnamefont
  {Langbehn}}, \bibinfo {author} {\bibfnamefont {Y.}~\bibnamefont {Peng}},
  \bibinfo {author} {\bibfnamefont {L.}~\bibnamefont {Trifunovic}}, \bibinfo
  {author} {\bibfnamefont {F.}~\bibnamefont {von Oppen}},\ and\ \bibinfo
  {author} {\bibfnamefont {P.~W.}\ \bibnamefont {Brouwer}},\ }\bibfield
  {title} {\bibinfo {title} {Reflection-symmetric second-order topological
  insulators and superconductors},\ }\href
  {https://doi.org/10.1103/PhysRevLett.119.246401} {\bibfield  {journal}
  {\bibinfo  {journal} {Phys. Rev. Lett.}\ }\textbf {\bibinfo {volume} {119}},\
  \bibinfo {pages} {246401} (\bibinfo {year} {2017})}\BibitemShut {NoStop}%
\bibitem [{\citenamefont {Zhu}(2018)}]{Zhu18}%
  \BibitemOpen
  \bibfield  {author} {\bibinfo {author} {\bibfnamefont {X.}~\bibnamefont
  {Zhu}},\ }\bibfield  {title} {\bibinfo {title} {Tunable majorana corner
  states in a two-dimensional second-order topological superconductor induced
  by magnetic fields},\ }\href {https://doi.org/10.1103/PhysRevB.97.205134}
  {\bibfield  {journal} {\bibinfo  {journal} {Phys. Rev. B}\ }\textbf {\bibinfo
  {volume} {97}},\ \bibinfo {pages} {205134} (\bibinfo {year}
  {2018})}\BibitemShut {NoStop}%
\bibitem [{\citenamefont {Khalaf}(2018)}]{Khalaf18}%
  \BibitemOpen
  \bibfield  {author} {\bibinfo {author} {\bibfnamefont {E.}~\bibnamefont
  {Khalaf}},\ }\bibfield  {title} {\bibinfo {title} {Higher-order topological
  insulators and superconductors protected by inversion symmetry},\ }\href
  {https://doi.org/10.1103/PhysRevB.97.205136} {\bibfield  {journal} {\bibinfo
  {journal} {Phys. Rev. B}\ }\textbf {\bibinfo {volume} {97}},\ \bibinfo
  {pages} {205136} (\bibinfo {year} {2018})}\BibitemShut {NoStop}%
\bibitem [{\citenamefont {Wang}\ \emph
  {et~al.}(2018{\natexlab{a}})\citenamefont {Wang}, \citenamefont {Lin},\ and\
  \citenamefont {Hughes}}]{Wang18}%
  \BibitemOpen
  \bibfield  {author} {\bibinfo {author} {\bibfnamefont {Y.}~\bibnamefont
  {Wang}}, \bibinfo {author} {\bibfnamefont {M.}~\bibnamefont {Lin}},\ and\
  \bibinfo {author} {\bibfnamefont {T.~L.}\ \bibnamefont {Hughes}},\ }\bibfield
   {title} {\bibinfo {title} {Weak-pairing higher order topological
  superconductors},\ }\href {https://doi.org/10.1103/PhysRevB.98.165144}
  {\bibfield  {journal} {\bibinfo  {journal} {Phys. Rev. B}\ }\textbf {\bibinfo
  {volume} {98}},\ \bibinfo {pages} {165144} (\bibinfo {year}
  {2018}{\natexlab{a}})}\BibitemShut {NoStop}%
\bibitem [{\citenamefont {Yan}\ \emph {et~al.}(2018)\citenamefont {Yan},
  \citenamefont {Song},\ and\ \citenamefont {Wang}}]{Yan18}%
  \BibitemOpen
  \bibfield  {author} {\bibinfo {author} {\bibfnamefont {Z.}~\bibnamefont
  {Yan}}, \bibinfo {author} {\bibfnamefont {F.}~\bibnamefont {Song}},\ and\
  \bibinfo {author} {\bibfnamefont {Z.}~\bibnamefont {Wang}},\ }\bibfield
  {title} {\bibinfo {title} {Majorana corner modes in a high-temperature
  platform},\ }\href {https://doi.org/10.1103/PhysRevLett.121.096803}
  {\bibfield  {journal} {\bibinfo  {journal} {Phys. Rev. Lett.}\ }\textbf
  {\bibinfo {volume} {121}},\ \bibinfo {pages} {096803} (\bibinfo {year}
  {2018})}\BibitemShut {NoStop}%
\bibitem [{\citenamefont {Wang}\ \emph
  {et~al.}(2018{\natexlab{b}})\citenamefont {Wang}, \citenamefont {Liu},
  \citenamefont {Lu},\ and\ \citenamefont {Zhang}}]{Fan18}%
  \BibitemOpen
  \bibfield  {author} {\bibinfo {author} {\bibfnamefont {Q.}~\bibnamefont
  {Wang}}, \bibinfo {author} {\bibfnamefont {C.-C.}\ \bibnamefont {Liu}},
  \bibinfo {author} {\bibfnamefont {Y.-M.}\ \bibnamefont {Lu}},\ and\ \bibinfo
  {author} {\bibfnamefont {F.}~\bibnamefont {Zhang}},\ }\bibfield  {title}
  {\bibinfo {title} {High-temperature majorana corner states},\ }\href
  {https://doi.org/10.1103/PhysRevLett.121.186801} {\bibfield  {journal}
  {\bibinfo  {journal} {Phys. Rev. Lett.}\ }\textbf {\bibinfo {volume} {121}},\
  \bibinfo {pages} {186801} (\bibinfo {year} {2018}{\natexlab{b}})}\BibitemShut
  {NoStop}%
\bibitem [{\citenamefont {Liu}\ \emph {et~al.}(2018)\citenamefont {Liu},
  \citenamefont {He},\ and\ \citenamefont {Nori}}]{Nori18}%
  \BibitemOpen
  \bibfield  {author} {\bibinfo {author} {\bibfnamefont {T.}~\bibnamefont
  {Liu}}, \bibinfo {author} {\bibfnamefont {J.~J.}\ \bibnamefont {He}},\ and\
  \bibinfo {author} {\bibfnamefont {F.}~\bibnamefont {Nori}},\ }\bibfield
  {title} {\bibinfo {title} {Majorana corner states in a two-dimensional
  magnetic topological insulator on a high-temperature superconductor},\ }\href
  {https://doi.org/10.1103/PhysRevB.98.245413} {\bibfield  {journal} {\bibinfo
  {journal} {Phys. Rev. B}\ }\textbf {\bibinfo {volume} {98}},\ \bibinfo
  {pages} {245413} (\bibinfo {year} {2018})}\BibitemShut {NoStop}%
\bibitem [{\citenamefont {Pan}\ \emph {et~al.}(2019)\citenamefont {Pan},
  \citenamefont {Yang}, \citenamefont {Chen}, \citenamefont {Xu}, \citenamefont
  {Liu},\ and\ \citenamefont {Liu}}]{Liu19}%
  \BibitemOpen
  \bibfield  {author} {\bibinfo {author} {\bibfnamefont {X.-H.}\ \bibnamefont
  {Pan}}, \bibinfo {author} {\bibfnamefont {K.-J.}\ \bibnamefont {Yang}},
  \bibinfo {author} {\bibfnamefont {L.}~\bibnamefont {Chen}}, \bibinfo {author}
  {\bibfnamefont {G.}~\bibnamefont {Xu}}, \bibinfo {author} {\bibfnamefont
  {C.-X.}\ \bibnamefont {Liu}},\ and\ \bibinfo {author} {\bibfnamefont
  {X.}~\bibnamefont {Liu}},\ }\bibfield  {title} {\bibinfo {title}
  {Lattice-symmetry-assisted second-order topological superconductors and
  majorana patterns},\ }\href {https://doi.org/10.1103/PhysRevLett.123.156801}
  {\bibfield  {journal} {\bibinfo  {journal} {Phys. Rev. Lett.}\ }\textbf
  {\bibinfo {volume} {123}},\ \bibinfo {pages} {156801} (\bibinfo {year}
  {2019})}\BibitemShut {NoStop}%
\bibitem [{\citenamefont {Hsu}\ \emph {et~al.}(2018)\citenamefont {Hsu},
  \citenamefont {Stano}, \citenamefont {Klinovaja},\ and\ \citenamefont
  {Loss}}]{Klin18}%
  \BibitemOpen
  \bibfield  {author} {\bibinfo {author} {\bibfnamefont {C.-H.}\ \bibnamefont
  {Hsu}}, \bibinfo {author} {\bibfnamefont {P.}~\bibnamefont {Stano}}, \bibinfo
  {author} {\bibfnamefont {J.}~\bibnamefont {Klinovaja}},\ and\ \bibinfo
  {author} {\bibfnamefont {D.}~\bibnamefont {Loss}},\ }\bibfield  {title}
  {\bibinfo {title} {Majorana kramers pairs in higher-order topological
  insulators},\ }\href {https://doi.org/10.1103/PhysRevLett.121.196801}
  {\bibfield  {journal} {\bibinfo  {journal} {Phys. Rev. Lett.}\ }\textbf
  {\bibinfo {volume} {121}},\ \bibinfo {pages} {196801} (\bibinfo {year}
  {2018})}\BibitemShut {NoStop}%
\bibitem [{\citenamefont {Volpez}\ \emph {et~al.}(2019)\citenamefont {Volpez},
  \citenamefont {Loss},\ and\ \citenamefont {Klinovaja}}]{Klin19}%
  \BibitemOpen
  \bibfield  {author} {\bibinfo {author} {\bibfnamefont {Y.}~\bibnamefont
  {Volpez}}, \bibinfo {author} {\bibfnamefont {D.}~\bibnamefont {Loss}},\ and\
  \bibinfo {author} {\bibfnamefont {J.}~\bibnamefont {Klinovaja}},\ }\bibfield
  {title} {\bibinfo {title} {Second-order topological superconductivity in
  $\ensuremath{\pi}$-junction rashba layers},\ }\href
  {https://doi.org/10.1103/PhysRevLett.122.126402} {\bibfield  {journal}
  {\bibinfo  {journal} {Phys. Rev. Lett.}\ }\textbf {\bibinfo {volume} {122}},\
  \bibinfo {pages} {126402} (\bibinfo {year} {2019})}\BibitemShut {NoStop}%
\bibitem [{\citenamefont {Zhu}(2019)}]{Xiaoyu19}%
  \BibitemOpen
  \bibfield  {author} {\bibinfo {author} {\bibfnamefont {X.}~\bibnamefont
  {Zhu}},\ }\bibfield  {title} {\bibinfo {title} {Second-order topological
  superconductors with mixed pairing},\ }\href
  {https://doi.org/10.1103/PhysRevLett.122.236401} {\bibfield  {journal}
  {\bibinfo  {journal} {Phys. Rev. Lett.}\ }\textbf {\bibinfo {volume} {122}},\
  \bibinfo {pages} {236401} (\bibinfo {year} {2019})}\BibitemShut {NoStop}%
\bibitem [{\citenamefont {Franca}\ \emph {et~al.}(2019)\citenamefont {Franca},
  \citenamefont {Efremov},\ and\ \citenamefont {Fulga}}]{Fulga19}%
  \BibitemOpen
  \bibfield  {author} {\bibinfo {author} {\bibfnamefont {S.}~\bibnamefont
  {Franca}}, \bibinfo {author} {\bibfnamefont {D.~V.}\ \bibnamefont
  {Efremov}},\ and\ \bibinfo {author} {\bibfnamefont {I.~C.}\ \bibnamefont
  {Fulga}},\ }\bibfield  {title} {\bibinfo {title} {Phase-tunable second-order
  topological superconductor},\ }\href
  {https://doi.org/10.1103/PhysRevB.100.075415} {\bibfield  {journal} {\bibinfo
   {journal} {Phys. Rev. B}\ }\textbf {\bibinfo {volume} {100}},\ \bibinfo
  {pages} {075415} (\bibinfo {year} {2019})}\BibitemShut {NoStop}%
\bibitem [{\citenamefont {Ghorashi}\ \emph {et~al.}(2019)\citenamefont
  {Ghorashi}, \citenamefont {Hu}, \citenamefont {Hughes},\ and\ \citenamefont
  {Rossi}}]{Ali19}%
  \BibitemOpen
  \bibfield  {author} {\bibinfo {author} {\bibfnamefont {S.~A.~A.}\
  \bibnamefont {Ghorashi}}, \bibinfo {author} {\bibfnamefont {X.}~\bibnamefont
  {Hu}}, \bibinfo {author} {\bibfnamefont {T.~L.}\ \bibnamefont {Hughes}},\
  and\ \bibinfo {author} {\bibfnamefont {E.}~\bibnamefont {Rossi}},\ }\bibfield
   {title} {\bibinfo {title} {Second-order dirac superconductors and magnetic
  field induced majorana hinge modes},\ }\href
  {https://doi.org/10.1103/PhysRevB.100.020509} {\bibfield  {journal} {\bibinfo
   {journal} {Phys. Rev. B}\ }\textbf {\bibinfo {volume} {100}},\ \bibinfo
  {pages} {020509} (\bibinfo {year} {2019})}\BibitemShut {NoStop}%
\bibitem [{\citenamefont {Zeng}\ \emph {et~al.}(2019)\citenamefont {Zeng},
  \citenamefont {Stanescu}, \citenamefont {Zhang}, \citenamefont {Scarola},\
  and\ \citenamefont {Tewari}}]{Zeng19}%
  \BibitemOpen
  \bibfield  {author} {\bibinfo {author} {\bibfnamefont {C.}~\bibnamefont
  {Zeng}}, \bibinfo {author} {\bibfnamefont {T.~D.}\ \bibnamefont {Stanescu}},
  \bibinfo {author} {\bibfnamefont {C.}~\bibnamefont {Zhang}}, \bibinfo
  {author} {\bibfnamefont {V.~W.}\ \bibnamefont {Scarola}},\ and\ \bibinfo
  {author} {\bibfnamefont {S.}~\bibnamefont {Tewari}},\ }\bibfield  {title}
  {\bibinfo {title} {Majorana corner modes with solitons in an attractive
  hubbard-hofstadter model of cold atom optical lattices},\ }\href
  {https://doi.org/10.1103/PhysRevLett.123.060402} {\bibfield  {journal}
  {\bibinfo  {journal} {Phys. Rev. Lett.}\ }\textbf {\bibinfo {volume} {123}},\
  \bibinfo {pages} {060402} (\bibinfo {year} {2019})}\BibitemShut {NoStop}%
\bibitem [{\citenamefont {Laubscher}\ \emph {et~al.}(2019)\citenamefont
  {Laubscher}, \citenamefont {Loss},\ and\ \citenamefont {Klinovaja}}]{Loss19}%
  \BibitemOpen
  \bibfield  {author} {\bibinfo {author} {\bibfnamefont {K.}~\bibnamefont
  {Laubscher}}, \bibinfo {author} {\bibfnamefont {D.}~\bibnamefont {Loss}},\
  and\ \bibinfo {author} {\bibfnamefont {J.}~\bibnamefont {Klinovaja}},\
  }\bibfield  {title} {\bibinfo {title} {Fractional topological
  superconductivity and parafermion corner states},\ }\href
  {https://doi.org/10.1103/PhysRevResearch.1.032017} {\bibfield  {journal}
  {\bibinfo  {journal} {Phys. Rev. Research}\ }\textbf {\bibinfo {volume}
  {1}},\ \bibinfo {pages} {032017} (\bibinfo {year} {2019})}\BibitemShut
  {NoStop}%
\bibitem [{\citenamefont {Roy}(2019)}]{RoB19}%
  \BibitemOpen
  \bibfield  {author} {\bibinfo {author} {\bibfnamefont {B.}~\bibnamefont
  {Roy}},\ }\bibfield  {title} {\bibinfo {title} {Antiunitary symmetry
  protected higher-order topological phases},\ }\href
  {https://doi.org/10.1103/PhysRevResearch.1.032048} {\bibfield  {journal}
  {\bibinfo  {journal} {Phys. Rev. Research}\ }\textbf {\bibinfo {volume}
  {1}},\ \bibinfo {pages} {032048} (\bibinfo {year} {2019})}\BibitemShut
  {NoStop}%
\bibitem [{\citenamefont {Zhang}\ and\ \citenamefont
  {Trauzettel}(2020)}]{Zhang20}%
  \BibitemOpen
  \bibfield  {author} {\bibinfo {author} {\bibfnamefont {S.-B.}\ \bibnamefont
  {Zhang}}\ and\ \bibinfo {author} {\bibfnamefont {B.}~\bibnamefont
  {Trauzettel}},\ }\bibfield  {title} {\bibinfo {title} {Detection of
  second-order topological superconductors by josephson junctions},\ }\href
  {https://doi.org/10.1103/PhysRevResearch.2.012018} {\bibfield  {journal}
  {\bibinfo  {journal} {Phys. Rev. Research}\ }\textbf {\bibinfo {volume}
  {2}},\ \bibinfo {pages} {012018} (\bibinfo {year} {2020})}\BibitemShut
  {NoStop}%
\bibitem [{\citenamefont {Roy}(2020)}]{Roy20}%
  \BibitemOpen
  \bibfield  {author} {\bibinfo {author} {\bibfnamefont {B.}~\bibnamefont
  {Roy}},\ }\bibfield  {title} {\bibinfo {title} {Higher-order topological
  superconductors in $\mathcal{P}$-, $\mathcal{T}$-odd quadrupolar dirac
  materials},\ }\href {https://doi.org/10.1103/PhysRevB.101.220506} {\bibfield
  {journal} {\bibinfo  {journal} {Phys. Rev. B}\ }\textbf {\bibinfo {volume}
  {101}},\ \bibinfo {pages} {220506} (\bibinfo {year} {2020})}\BibitemShut
  {NoStop}%
\bibitem [{\citenamefont {Wu}\ \emph {et~al.}(2020)\citenamefont {Wu},
  \citenamefont {Hou}, \citenamefont {Li}, \citenamefont {Luo}, \citenamefont
  {Shi},\ and\ \citenamefont {Zhang}}]{Chuan20}%
  \BibitemOpen
  \bibfield  {author} {\bibinfo {author} {\bibfnamefont {Y.-J.}\ \bibnamefont
  {Wu}}, \bibinfo {author} {\bibfnamefont {J.}~\bibnamefont {Hou}}, \bibinfo
  {author} {\bibfnamefont {Y.-M.}\ \bibnamefont {Li}}, \bibinfo {author}
  {\bibfnamefont {X.-W.}\ \bibnamefont {Luo}}, \bibinfo {author} {\bibfnamefont
  {X.}~\bibnamefont {Shi}},\ and\ \bibinfo {author} {\bibfnamefont
  {C.}~\bibnamefont {Zhang}},\ }\bibfield  {title} {\bibinfo {title} {In-plane
  zeeman-field-induced majorana corner and hinge modes in an $s$-wave
  superconductor heterostructure},\ }\href
  {https://doi.org/10.1103/PhysRevLett.124.227001} {\bibfield  {journal}
  {\bibinfo  {journal} {Phys. Rev. Lett.}\ }\textbf {\bibinfo {volume} {124}},\
  \bibinfo {pages} {227001} (\bibinfo {year} {2020})}\BibitemShut {NoStop}%
\bibitem [{\citenamefont {Kheirkhah}\ \emph {et~al.}(2020)\citenamefont
  {Kheirkhah}, \citenamefont {Yan}, \citenamefont {Nagai},\ and\ \citenamefont
  {Marsiglio}}]{Nagai20}%
  \BibitemOpen
  \bibfield  {author} {\bibinfo {author} {\bibfnamefont {M.}~\bibnamefont
  {Kheirkhah}}, \bibinfo {author} {\bibfnamefont {Z.}~\bibnamefont {Yan}},
  \bibinfo {author} {\bibfnamefont {Y.}~\bibnamefont {Nagai}},\ and\ \bibinfo
  {author} {\bibfnamefont {F.}~\bibnamefont {Marsiglio}},\ }\bibfield  {title}
  {\bibinfo {title} {First- and second-order topological superconductivity and
  temperature-driven topological phase transitions in the extended hubbard
  model with spin-orbit coupling},\ }\href
  {https://doi.org/10.1103/PhysRevLett.125.017001} {\bibfield  {journal}
  {\bibinfo  {journal} {Phys. Rev. Lett.}\ }\textbf {\bibinfo {volume} {125}},\
  \bibinfo {pages} {017001} (\bibinfo {year} {2020})}\BibitemShut {NoStop}%
\bibitem [{\citenamefont {Zhang}\ \emph
  {et~al.}(2020{\natexlab{a}})\citenamefont {Zhang}, \citenamefont {Calzona},\
  and\ \citenamefont {Trauzettel}}]{Zhang201}%
  \BibitemOpen
  \bibfield  {author} {\bibinfo {author} {\bibfnamefont {S.-B.}\ \bibnamefont
  {Zhang}}, \bibinfo {author} {\bibfnamefont {A.}~\bibnamefont {Calzona}},\
  and\ \bibinfo {author} {\bibfnamefont {B.}~\bibnamefont {Trauzettel}},\
  }\bibfield  {title} {\bibinfo {title} {All-electrically tunable networks of
  majorana bound states},\ }\href {https://doi.org/10.1103/PhysRevB.102.100503}
  {\bibfield  {journal} {\bibinfo  {journal} {Phys. Rev. B}\ }\textbf {\bibinfo
  {volume} {102}},\ \bibinfo {pages} {100503} (\bibinfo {year}
  {2020}{\natexlab{a}})}\BibitemShut {NoStop}%
\bibitem [{\citenamefont {Zhang}\ \emph
  {et~al.}(2020{\natexlab{b}})\citenamefont {Zhang}, \citenamefont {Rui},
  \citenamefont {Calzona}, \citenamefont {Choi}, \citenamefont {Schnyder},\
  and\ \citenamefont {Trauzettel}}]{Zhang202}%
  \BibitemOpen
  \bibfield  {author} {\bibinfo {author} {\bibfnamefont {S.-B.}\ \bibnamefont
  {Zhang}}, \bibinfo {author} {\bibfnamefont {W.~B.}\ \bibnamefont {Rui}},
  \bibinfo {author} {\bibfnamefont {A.}~\bibnamefont {Calzona}}, \bibinfo
  {author} {\bibfnamefont {S.-J.}\ \bibnamefont {Choi}}, \bibinfo {author}
  {\bibfnamefont {A.~P.}\ \bibnamefont {Schnyder}},\ and\ \bibinfo {author}
  {\bibfnamefont {B.}~\bibnamefont {Trauzettel}},\ }\bibfield  {title}
  {\bibinfo {title} {Topological and holonomic quantum computation based on
  second-order topological superconductors},\ }\href
  {https://doi.org/10.1103/PhysRevResearch.2.043025} {\bibfield  {journal}
  {\bibinfo  {journal} {Phys. Rev. Research}\ }\textbf {\bibinfo {volume}
  {2}},\ \bibinfo {pages} {043025} (\bibinfo {year}
  {2020}{\natexlab{b}})}\BibitemShut {NoStop}%
\bibitem [{\citenamefont {Li}\ and\ \citenamefont {Zhou}(2021)}]{Tao21}%
  \BibitemOpen
  \bibfield  {author} {\bibinfo {author} {\bibfnamefont {Y.-X.}\ \bibnamefont
  {Li}}\ and\ \bibinfo {author} {\bibfnamefont {T.}~\bibnamefont {Zhou}},\
  }\bibfield  {title} {\bibinfo {title} {Rotational symmetry breaking and
  partial majorana corner states in a heterostructure based on high-${T}_{c}$
  superconductors},\ }\href {https://doi.org/10.1103/PhysRevB.103.024517}
  {\bibfield  {journal} {\bibinfo  {journal} {Phys. Rev. B}\ }\textbf {\bibinfo
  {volume} {103}},\ \bibinfo {pages} {024517} (\bibinfo {year}
  {2021})}\BibitemShut {NoStop}%
\bibitem [{\citenamefont {Kheirkhah}\ \emph {et~al.}(2021)\citenamefont
  {Kheirkhah}, \citenamefont {Yan},\ and\ \citenamefont {Marsiglio}}]{Frank21}%
  \BibitemOpen
  \bibfield  {author} {\bibinfo {author} {\bibfnamefont {M.}~\bibnamefont
  {Kheirkhah}}, \bibinfo {author} {\bibfnamefont {Z.}~\bibnamefont {Yan}},\
  and\ \bibinfo {author} {\bibfnamefont {F.}~\bibnamefont {Marsiglio}},\
  }\bibfield  {title} {\bibinfo {title} {Vortex-line topology in iron-based
  superconductors with and without second-order topology},\ }\href
  {https://doi.org/10.1103/PhysRevB.103.L140502} {\bibfield  {journal}
  {\bibinfo  {journal} {Phys. Rev. B}\ }\textbf {\bibinfo {volume} {103}},\
  \bibinfo {pages} {L140502} (\bibinfo {year} {2021})}\BibitemShut {NoStop}%
\bibitem [{\citenamefont {Luo}\ \emph {et~al.}(2021)\citenamefont {Luo},
  \citenamefont {Pan},\ and\ \citenamefont {Liu}}]{Xin21}%
  \BibitemOpen
  \bibfield  {author} {\bibinfo {author} {\bibfnamefont {X.-J.}\ \bibnamefont
  {Luo}}, \bibinfo {author} {\bibfnamefont {X.-H.}\ \bibnamefont {Pan}},\ and\
  \bibinfo {author} {\bibfnamefont {X.}~\bibnamefont {Liu}},\ }\bibfield
  {title} {\bibinfo {title} {Higher-order topological superconductors based on
  weak topological insulators},\ }\href
  {https://doi.org/10.1103/PhysRevB.104.104510} {\bibfield  {journal} {\bibinfo
   {journal} {Phys. Rev. B}\ }\textbf {\bibinfo {volume} {104}},\ \bibinfo
  {pages} {104510} (\bibinfo {year} {2021})}\BibitemShut {NoStop}%
\bibitem [{\citenamefont {Ren}\ \emph {et~al.}(2020)\citenamefont {Ren},
  \citenamefont {Qiao},\ and\ \citenamefont {Niu}}]{Niu20}%
  \BibitemOpen
  \bibfield  {author} {\bibinfo {author} {\bibfnamefont {Y.}~\bibnamefont
  {Ren}}, \bibinfo {author} {\bibfnamefont {Z.}~\bibnamefont {Qiao}},\ and\
  \bibinfo {author} {\bibfnamefont {Q.}~\bibnamefont {Niu}},\ }\bibfield
  {title} {\bibinfo {title} {Engineering corner states from two-dimensional
  topological insulators},\ }\href
  {https://doi.org/10.1103/PhysRevLett.124.166804} {\bibfield  {journal}
  {\bibinfo  {journal} {Phys. Rev. Lett.}\ }\textbf {\bibinfo {volume} {124}},\
  \bibinfo {pages} {166804} (\bibinfo {year} {2020})}\BibitemShut {NoStop}%
\bibitem [{\citenamefont {Shen}(2017)}]{shenbook}%
  \BibitemOpen
  \bibfield  {author} {\bibinfo {author} {\bibfnamefont {S.-Q.}\ \bibnamefont
  {Shen}},\ }\bibfield  {title} {\bibinfo {title} {Topological insulators:
  Dirac equation in condensed matter, 2nd ed.},\ }\href@noop {} {\bibfield
  {journal} {\bibinfo  {journal} {(Springer, Singapore)}\ } (\bibinfo {year}
  {2017})}\BibitemShut {NoStop}%
\bibitem [{\citenamefont {Jackiw}\ and\ \citenamefont {Rebbi}(1976)}]{JR}%
  \BibitemOpen
  \bibfield  {author} {\bibinfo {author} {\bibfnamefont {R.}~\bibnamefont
  {Jackiw}}\ and\ \bibinfo {author} {\bibfnamefont {C.}~\bibnamefont {Rebbi}},\
  }\bibfield  {title} {\bibinfo {title} {Solitons with fermion number
  \textonehalf{}},\ }\href {https://doi.org/10.1103/PhysRevD.13.3398}
  {\bibfield  {journal} {\bibinfo  {journal} {Phys. Rev. D}\ }\textbf {\bibinfo
  {volume} {13}},\ \bibinfo {pages} {3398} (\bibinfo {year}
  {1976})}\BibitemShut {NoStop}%
\bibitem [{\citenamefont {Modak}\ \emph {et~al.}(2012)\citenamefont {Modak},
  \citenamefont {Sengupta},\ and\ \citenamefont {Sen}}]{Modak12}%
  \BibitemOpen
  \bibfield  {author} {\bibinfo {author} {\bibfnamefont {S.}~\bibnamefont
  {Modak}}, \bibinfo {author} {\bibfnamefont {K.}~\bibnamefont {Sengupta}},\
  and\ \bibinfo {author} {\bibfnamefont {D.}~\bibnamefont {Sen}},\ }\bibfield
  {title} {\bibinfo {title} {Spin injection into a metal from a topological
  insulator},\ }\href {https://doi.org/10.1103/PhysRevB.86.205114} {\bibfield
  {journal} {\bibinfo  {journal} {Phys. Rev. B}\ }\textbf {\bibinfo {volume}
  {86}},\ \bibinfo {pages} {205114} (\bibinfo {year} {2012})}\BibitemShut
  {NoStop}%
\bibitem [{\citenamefont {Soori}\ \emph {et~al.}(2013)\citenamefont {Soori},
  \citenamefont {Deb}, \citenamefont {Sengupta},\ and\ \citenamefont
  {Sen}}]{Soori13}%
  \BibitemOpen
  \bibfield  {author} {\bibinfo {author} {\bibfnamefont {A.}~\bibnamefont
  {Soori}}, \bibinfo {author} {\bibfnamefont {O.}~\bibnamefont {Deb}}, \bibinfo
  {author} {\bibfnamefont {K.}~\bibnamefont {Sengupta}},\ and\ \bibinfo
  {author} {\bibfnamefont {D.}~\bibnamefont {Sen}},\ }\bibfield  {title}
  {\bibinfo {title} {Transport across a junction of topological insulators and
  a superconductor},\ }\href {https://doi.org/10.1103/PhysRevB.87.245435}
  {\bibfield  {journal} {\bibinfo  {journal} {Phys. Rev. B}\ }\textbf {\bibinfo
  {volume} {87}},\ \bibinfo {pages} {245435} (\bibinfo {year}
  {2013})}\BibitemShut {NoStop}%
\bibitem [{\citenamefont {Soori}(2020)}]{Soori20}%
  \BibitemOpen
  \bibfield  {author} {\bibinfo {author} {\bibfnamefont {A.}~\bibnamefont
  {Soori}},\ }\href@noop {} {\bibinfo {title} {Scattering in quantum wires and
  junctions of quantum wires with edge states of quantum spin hall insulators}}
  (\bibinfo {year} {2020}),\ \Eprint {https://arxiv.org/abs/2005.11557}
  {arXiv:2005.11557 [cond-mat.mes-hall]} \BibitemShut {NoStop}%
\bibitem [{\citenamefont {Blonder}\ \emph {et~al.}(1982)\citenamefont
  {Blonder}, \citenamefont {Tinkham},\ and\ \citenamefont {Klapwijk}}]{BTK}%
  \BibitemOpen
  \bibfield  {author} {\bibinfo {author} {\bibfnamefont {G.~E.}\ \bibnamefont
  {Blonder}}, \bibinfo {author} {\bibfnamefont {M.}~\bibnamefont {Tinkham}},\
  and\ \bibinfo {author} {\bibfnamefont {T.~M.}\ \bibnamefont {Klapwijk}},\
  }\bibfield  {title} {\bibinfo {title} {Transition from metallic to tunneling
  regimes in superconducting microconstrictions: Excess current, charge
  imbalance, and supercurrent conversion},\ }\href
  {https://doi.org/10.1103/PhysRevB.25.4515} {\bibfield  {journal} {\bibinfo
  {journal} {Phys. Rev. B}\ }\textbf {\bibinfo {volume} {25}},\ \bibinfo
  {pages} {4515} (\bibinfo {year} {1982})}\BibitemShut {NoStop}%
\bibitem [{\citenamefont {Sengupta}\ \emph {et~al.}(2001)\citenamefont
  {Sengupta}, \citenamefont {\ifmmode \check{Z}\else
  \v{Z}\fi{}uti\ifmmode~\acute{c}\else \'{c}\fi{}}, \citenamefont {Kwon},
  \citenamefont {Yakovenko},\ and\ \citenamefont {Das~Sarma}}]{Sengupta01}%
  \BibitemOpen
  \bibfield  {author} {\bibinfo {author} {\bibfnamefont {K.}~\bibnamefont
  {Sengupta}}, \bibinfo {author} {\bibfnamefont {I.}~\bibnamefont {\ifmmode
  \check{Z}\else \v{Z}\fi{}uti\ifmmode~\acute{c}\else \'{c}\fi{}}}, \bibinfo
  {author} {\bibfnamefont {H.-J.}\ \bibnamefont {Kwon}}, \bibinfo {author}
  {\bibfnamefont {V.~M.}\ \bibnamefont {Yakovenko}},\ and\ \bibinfo {author}
  {\bibfnamefont {S.}~\bibnamefont {Das~Sarma}},\ }\bibfield  {title} {\bibinfo
  {title} {Midgap edge states and pairing symmetry of quasi-one-dimensional
  organic superconductors},\ }\href
  {https://doi.org/10.1103/PhysRevB.63.144531} {\bibfield  {journal} {\bibinfo
  {journal} {Phys. Rev. B}\ }\textbf {\bibinfo {volume} {63}},\ \bibinfo
  {pages} {144531} (\bibinfo {year} {2001})}\BibitemShut {NoStop}%
\bibitem [{\citenamefont {Bolech}\ and\ \citenamefont
  {Demler}(2007)}]{Bolech07}%
  \BibitemOpen
  \bibfield  {author} {\bibinfo {author} {\bibfnamefont {C.~J.}\ \bibnamefont
  {Bolech}}\ and\ \bibinfo {author} {\bibfnamefont {E.}~\bibnamefont
  {Demler}},\ }\bibfield  {title} {\bibinfo {title} {Observing majorana bound
  states in $p$-wave superconductors using noise measurements in tunneling
  experiments},\ }\href {https://doi.org/10.1103/PhysRevLett.98.237002}
  {\bibfield  {journal} {\bibinfo  {journal} {Phys. Rev. Lett.}\ }\textbf
  {\bibinfo {volume} {98}},\ \bibinfo {pages} {237002} (\bibinfo {year}
  {2007})}\BibitemShut {NoStop}%
\bibitem [{\citenamefont {Akhmerov}\ \emph {et~al.}(2009)\citenamefont
  {Akhmerov}, \citenamefont {Nilsson},\ and\ \citenamefont
  {Beenakker}}]{Akhmerov09}%
  \BibitemOpen
  \bibfield  {author} {\bibinfo {author} {\bibfnamefont {A.~R.}\ \bibnamefont
  {Akhmerov}}, \bibinfo {author} {\bibfnamefont {J.}~\bibnamefont {Nilsson}},\
  and\ \bibinfo {author} {\bibfnamefont {C.~W.~J.}\ \bibnamefont {Beenakker}},\
  }\bibfield  {title} {\bibinfo {title} {Electrically detected interferometry
  of majorana fermions in a topological insulator},\ }\href
  {https://doi.org/10.1103/PhysRevLett.102.216404} {\bibfield  {journal}
  {\bibinfo  {journal} {Phys. Rev. Lett.}\ }\textbf {\bibinfo {volume} {102}},\
  \bibinfo {pages} {216404} (\bibinfo {year} {2009})}\BibitemShut {NoStop}%
\bibitem [{\citenamefont {Tanaka}\ \emph {et~al.}(2009)\citenamefont {Tanaka},
  \citenamefont {Yokoyama},\ and\ \citenamefont {Nagaosa}}]{Tanaka09}%
  \BibitemOpen
  \bibfield  {author} {\bibinfo {author} {\bibfnamefont {Y.}~\bibnamefont
  {Tanaka}}, \bibinfo {author} {\bibfnamefont {T.}~\bibnamefont {Yokoyama}},\
  and\ \bibinfo {author} {\bibfnamefont {N.}~\bibnamefont {Nagaosa}},\
  }\bibfield  {title} {\bibinfo {title} {Manipulation of the majorana fermion,
  andreev reflection, and josephson current on topological insulators},\ }\href
  {https://doi.org/10.1103/PhysRevLett.103.107002} {\bibfield  {journal}
  {\bibinfo  {journal} {Phys. Rev. Lett.}\ }\textbf {\bibinfo {volume} {103}},\
  \bibinfo {pages} {107002} (\bibinfo {year} {2009})}\BibitemShut {NoStop}%
\bibitem [{\citenamefont {Law}\ \emph {et~al.}(2009)\citenamefont {Law},
  \citenamefont {Lee},\ and\ \citenamefont {Ng}}]{Law09}%
  \BibitemOpen
  \bibfield  {author} {\bibinfo {author} {\bibfnamefont {K.~T.}\ \bibnamefont
  {Law}}, \bibinfo {author} {\bibfnamefont {P.~A.}\ \bibnamefont {Lee}},\ and\
  \bibinfo {author} {\bibfnamefont {T.~K.}\ \bibnamefont {Ng}},\ }\bibfield
  {title} {\bibinfo {title} {Majorana fermion induced resonant andreev
  reflection},\ }\href {https://doi.org/10.1103/PhysRevLett.103.237001}
  {\bibfield  {journal} {\bibinfo  {journal} {Phys. Rev. Lett.}\ }\textbf
  {\bibinfo {volume} {103}},\ \bibinfo {pages} {237001} (\bibinfo {year}
  {2009})}\BibitemShut {NoStop}%
\bibitem [{\citenamefont {Bocquillon}\ \emph {et~al.}(2017)\citenamefont
  {Bocquillon}, \citenamefont {Deacon}, \citenamefont {Wiedenmann},
  \citenamefont {Leubner}, \citenamefont {Klapwijk}, \citenamefont {Br{\"u}ne},
  \citenamefont {Ishibashi}, \citenamefont {Buhmann},\ and\ \citenamefont
  {Molenkamp}}]{Bocquillon17}%
  \BibitemOpen
  \bibfield  {author} {\bibinfo {author} {\bibfnamefont {E.}~\bibnamefont
  {Bocquillon}}, \bibinfo {author} {\bibfnamefont {R.~S.}\ \bibnamefont
  {Deacon}}, \bibinfo {author} {\bibfnamefont {J.}~\bibnamefont {Wiedenmann}},
  \bibinfo {author} {\bibfnamefont {P.}~\bibnamefont {Leubner}}, \bibinfo
  {author} {\bibfnamefont {T.~M.}\ \bibnamefont {Klapwijk}}, \bibinfo {author}
  {\bibfnamefont {C.}~\bibnamefont {Br{\"u}ne}}, \bibinfo {author}
  {\bibfnamefont {K.}~\bibnamefont {Ishibashi}}, \bibinfo {author}
  {\bibfnamefont {H.}~\bibnamefont {Buhmann}},\ and\ \bibinfo {author}
  {\bibfnamefont {L.~W.}\ \bibnamefont {Molenkamp}},\ }\bibfield  {title}
  {\bibinfo {title} {{Gapless Andreev bound states in the quantum spin Hall
  insulator HgTe}},\ }\href {https://doi.org/10.1038/nnano.2016.159} {\bibfield
   {journal} {\bibinfo  {journal} {Nature Nanotechnology}\ }\textbf {\bibinfo
  {volume} {12}},\ \bibinfo {pages} {137} (\bibinfo {year} {2017})}\BibitemShut
  {NoStop}%
\bibitem [{\citenamefont {Liu}\ \emph {et~al.}(2013)\citenamefont {Liu},
  \citenamefont {Hsu},\ and\ \citenamefont {Liu}}]{LiuX2013}%
  \BibitemOpen
  \bibfield  {author} {\bibinfo {author} {\bibfnamefont {X.}~\bibnamefont
  {Liu}}, \bibinfo {author} {\bibfnamefont {H.-C.}\ \bibnamefont {Hsu}},\ and\
  \bibinfo {author} {\bibfnamefont {C.-X.}\ \bibnamefont {Liu}},\ }\bibfield
  {title} {\bibinfo {title} {In-plane magnetization-induced quantum anomalous
  hall effect},\ }\href {https://doi.org/10.1103/PhysRevLett.111.086802}
  {\bibfield  {journal} {\bibinfo  {journal} {Phys. Rev. Lett.}\ }\textbf
  {\bibinfo {volume} {111}},\ \bibinfo {pages} {086802} (\bibinfo {year}
  {2013})}\BibitemShut {NoStop}%
\bibitem [{\citenamefont {Ren}\ \emph {et~al.}(2019)\citenamefont {Ren},
  \citenamefont {Pientka}, \citenamefont {Hart}, \citenamefont {Pierce},
  \citenamefont {Kosowsky}, \citenamefont {Lunczer}, \citenamefont {Schlereth},
  \citenamefont {Scharf}, \citenamefont {Hankiewicz}, \citenamefont
  {Molenkamp}, \citenamefont {Halperin},\ and\ \citenamefont
  {Yacoby}}]{Ren2019}%
  \BibitemOpen
  \bibfield  {author} {\bibinfo {author} {\bibfnamefont {H.}~\bibnamefont
  {Ren}}, \bibinfo {author} {\bibfnamefont {F.}~\bibnamefont {Pientka}},
  \bibinfo {author} {\bibfnamefont {S.}~\bibnamefont {Hart}}, \bibinfo {author}
  {\bibfnamefont {A.~T.}\ \bibnamefont {Pierce}}, \bibinfo {author}
  {\bibfnamefont {M.}~\bibnamefont {Kosowsky}}, \bibinfo {author}
  {\bibfnamefont {L.}~\bibnamefont {Lunczer}}, \bibinfo {author} {\bibfnamefont
  {R.}~\bibnamefont {Schlereth}}, \bibinfo {author} {\bibfnamefont
  {B.}~\bibnamefont {Scharf}}, \bibinfo {author} {\bibfnamefont {E.~M.}\
  \bibnamefont {Hankiewicz}}, \bibinfo {author} {\bibfnamefont {L.~W.}\
  \bibnamefont {Molenkamp}}, \bibinfo {author} {\bibfnamefont {B.~I.}\
  \bibnamefont {Halperin}},\ and\ \bibinfo {author} {\bibfnamefont
  {A.}~\bibnamefont {Yacoby}},\ }\bibfield  {title} {\bibinfo {title}
  {Topological superconductivity in a phase-controlled josephson junction},\
  }\href {https://doi.org/10.1038/s41586-019-1148-9} {\bibfield  {journal}
  {\bibinfo  {journal} {Nature}\ }\textbf {\bibinfo {volume} {569}},\ \bibinfo
  {pages} {93} (\bibinfo {year} {2019})}\BibitemShut {NoStop}%
\end{thebibliography}%
		
\end{document}